\documentclass[10pt,twocolumn,amsmath,amssymb,aps,prb,showpacs,longbibliography,superscriptaddress]{revtex4-1}
\usepackage[latin9]{inputenc}
\usepackage{amsmath}
\usepackage{amssymb}
\usepackage{graphicx}
\usepackage{wasysym}
\usepackage{esint}

\makeatletter

\usepackage{amsfonts}
\usepackage{graphics}

\makeatother

\begin{document}
\title{Finite-momentum superconductivity with singlet-triplet mixing in an
altermagnetic metal: A pairing instability analysis}
\author{Hui Hu}
\email{hhu@swin.edu.au}

\affiliation{Centre for Quantum Technology Theory, Swinburne University of Technology,
Melbourne 3122, Australia}
\author{Zhao Liu}
\affiliation{Centre for Quantum Technology Theory, Swinburne University of Technology,
Melbourne 3122, Australia}
\author{Jia Wang}
\affiliation{Centre for Quantum Technology Theory, Swinburne University of Technology,
Melbourne 3122, Australia}
\author{Xia-Ji Liu}
\affiliation{Centre for Quantum Technology Theory, Swinburne University of Technology,
Melbourne 3122, Australia}
\author{Yoji Ohashi}
\email{yohashi@keio.jp}

\affiliation{Department of Physics, Keio University, Yokohama, Kanagawa 223-8522,
Japan}
\date{\today}
\begin{abstract}
We analyze the pairing instability of an altermagnetic metal on a
square lattice driven by an attractive nearest-neighbor interaction.
This interaction enables multiple pairing channels, including even-parity
extended $s$-wave and $d$-wave states, as well as two odd-parity
$p$-wave channels. We verify that altermagnetic spin-splitting in
the single-particle dispersion gives rise to finite-momentum pairing
between electrons with unlike spins, in agreement with earlier predictions.
Quite unexpectedly, this pairing typically emerges across multiple
channels with mixed parity. Consequently, the resulting finite-momentum
Fulde-Ferrell-Larkin-Ovchinnikov (FFLO) superconducting phase is expected
to exhibit a multi-component order parameter featuring singlet-triplet
mixing. We examine several forms of altermagnetism, specifically $d_{xy}$-wave
and $d_{x^{2}-y^{2}}$-wave altermagnetic couplings, and present the
corresponding phase diagrams. Additionally, we study triplet pairing
between electrons with identical spins and find that it always occurs
at zero center-of-mass momentum. Although it is unfavorable in the
regime of weak altermagnetic coupling and low electron filling, it
can dominate the phase diagram when the altermagnetic coupling is
sufficiently strong. The influence of on-site attractive interactions
on mixed-parity pairing is also explored.
\end{abstract}
\maketitle

\section{Introduction}

A newly identified magnetic order, altermagnetism, has rapidly emerged
as a subject of intense interest \citep{Smejkal2022,Bai2024,Jungwirth2025,Liu2025Review}.
It transcends the conventional classification into ferromagnetic and
antiferromagnetic phase by combining zero net magnetization with a
momentum-dependent spin splitting of electronic bands \citep{Noda2016,Naka2019,Hayami2019,Ahn2019,Hayami2020,Smejkal2020,Mazin2021,Mazin2023,McClarty2024,Roig2024}.
This spin polarization is enforced by crystalline symmetries rather
than relativistic spin-orbit coupling \citep{Smejkal2022,Xiao2024,Chen2024,JiangY2024},
leading to spin-resolved Fermi surfaces in the absence of macroscopic
magnetic fields. As a result, altermagnets occupy a unique position
among magnetically ordered quantum materials \citep{Smejkal2022,Bai2024,Jungwirth2025,Liu2025Review}.
Experimental signatures of altermagnetism have now been reported in
an increasing number of compounds \citep{Krempasky2024,Amin2024,Jiang2025,Zhang2025},
underscoring its relevance as a generic and robust form of magnetic
order.

Spin-split Fermi surfaces in the absence of net magnetization have
profound implications for quantum many-body phenomena \citep{Jungwirth2025,Liu2025Review,Mazin2025,Fukaya2025Review},
including superconductivity \citep{Ouassou2023,Sumita2023,Sun2023,Zhu2023,Brekke2023,Lu2024,Zhang2024,Chakraborty2024,Maeland2024,Cheng2024,Bose2024,deCarvalho2024,Chakraborty2025,Mukasa2025,Sim2025,Hong2025,Fukaya2025,Sumita2025,Hu2025PRB,Iorsh2025,Hu2025AB,Liu2026PRB,Jasiewicz2026,Fu2026,Liu2026,Hu2026,Monkman2026,Zhao2026}.
The intrinsic spin splitting associated with magnetic order strongly
influences Cooper pairing and can modify the stability of different
superconducting states. In conventional ferromagnets, the exchange
field generates a uniform spin splitting that suppresses zero-momentum
singlet pairing and, under suitable conditions, gives rise to the
Fulde-Ferrell-Larkin-Ovchinnikov (FFLO) state through pairing between
electrons with finite center-of-mass momentum \citep{Fulde1964,Larkin1964,Casalbuoni2004,Hu2006,Hu2007,Liu2007,Radzihovsky2010,Gubbels2013,Liu2013,Cao2014,Sheehy2015,Wang2018,Kawamura2022,Kawamura2024}.
However, such states generally require strong magnetic fields and
remain fragile because of orbital depairing and competing electronic
instabilities \citep{Uji2006,Liao2010,Zhao2023}. In contrast, conventional
antiferromagnets preserve combined parity-time $\mathcal{PT}$ symmetry,
which protects spin degeneracy and therefore generally favors zero-momentum
BCS pairing. Altermagnets provide a fundamentally different route
to finite-momentum superconductivity \citep{Sumita2023,Zhang2024,Chakraborty2024,Chakraborty2025,Mukasa2025,Sim2025,Hong2025,Sumita2025,Hu2025PRB,Iorsh2025,Hu2025AB,Liu2026PRB,Jasiewicz2026}.
Their crystal-symmetry-enforced, momentum-dependent spin splitting
breaks spin degeneracy without producing a net magnetization, yielding
anisotropic spin textures across the Brillouin zone. Consequently,
finite-momentum pairing may emerge intrinsically by connecting symmetry-related
regions of the Fermi surface with compatible spin polarization, rather
than by compensating a uniform exchange splitting \citep{Soto-Garrido2014}.
Because this mechanism is dictated by the underlying crystal symmetry
instead of external magnetic fields, it avoids orbital pair breaking
and directly links the pairing wave vector to the electronic structure.
As a result, altermagnets provide a promising platform for realizing
robust, highly anisotropic superconducting phases, including finite-momentum
states such as pair-density-wave and multi-$Q$ phases \citep{Sumita2025},
and for exploring how magnetic symmetry influences superconductivity
\citep{Liu2025Review}.

To date, altermagnetism-induced FFLO superconductivity has been explored
primarily through theoretical studies based on two-dimensional lattice
Hamiltonians with either on-site \citep{Mukasa2025,Hong2025,Sumita2025,Liu2026}
or nearest-neighbor \citep{Chakraborty2024,Chakraborty2025,Hong2025,Jasiewicz2026}
attractive interactions. The on-site attraction case is comparatively
straightforward, as it supports a single $s$-wave pairing channel.
In this setting, the physical mechanism underlying FFLO formation
is most transparently revealed within a low-filling continuum description
\citep{Zhang2024,Hu2025PRB,Hu2025AB,Liu2026}: sufficiently strong
altermagnetic coupling induces a continuous transition from conventional
Bardeen-Cooper-Schrieffer (BCS) pairing to an inhomogeneous FFLO state,
giving rise to rich phase diagrams under magnetic field featuring
quantum Lifshitz multi-critical points separating polarized BCS, FFLO,
and normal phases. 

By contrast, nearest-neighbor attraction presents a more intricate
situation, since it supports multiple competing pairing channels \citep{Zhu2023,Hong2025,Hu2026},
including extended $s$-wave, two $p$-wave, and $d$-wave pairings.
In the absence of spin-orbit coupling, most mean-field studies restrict
attention to a single dominant channel, assuming either spin-singlet
($s$- or $d$-wave) or spin-triplet ($p$-wave) pairing. Early works
focused on topological $p$-wave superconductivity and its competition
with $s$-wave order \citep{Zhu2023}, followed by investigations
of $d$-wave FFLO superconductivity with and without external magnetic
fields \citep{Chakraborty2024}. A notable exception is the recent
work by Jasiewicz \textit{et al.} \citep{Jasiewicz2026}, who demonstrated
singlet-triplet mixing with coexisting $d$-wave and $p$-wave FFLO
pairings induced by $d_{x^{2}-y^{2}}$-type altermagnetism at relatively
high electron filling. This result is further corroborated by an exact
two-electron analysis \citep{Hu2026}, showing that altermagnetism
stabilizes bound pairs with finite center-of-mass momentum, whose
ground state consists of a coherent superposition of spin-singlet
and spin-triplet components.

In this work, we present a systematic study of singlet-triplet mixing
in Cooper pairing within altermagnetic metals over a wide range of
electron fillings and for different altermagnetic symmetries. To circumvent
the numerical complexity associated with solving for multicomponent
superconducting order parameters, we employ a pairing-instability
analysis for two electrons near the spin-split Fermi surfaces based
on the Thouless criterion \citep{Thouless1960,Liu2006}. This approach
naturally extends the earlier exact two-body calculations \citep{Hu2026}
by incorporating many-body effects arising from the presence of a
Fermi sea \citep{Cooper1956}. We find that the leading pairing instability
generically occurs at finite center-of-mass momentum and simultaneously
involves spin-singlet and spin-triplet channels with mixed parity.
Consequently, the superconducting phase is anticipated to host an
FFLO state described by a multicomponent order parameter with intrinsic
singlet-triplet mixing \citep{Jasiewicz2026}. We determine the superconducting
phase diagrams and quantify the relative weights of the distinct pairing
channels at the superconducting transition, as functions of electron
filling and interaction strength for various types of altermagnetism.

The rest of this paper is organized as follows. In Sec. II, we introduce
the model Hamiltonian. Section III formulates the many-body $T$-matrix
approximation for the two-particle vertex function in the Cooper pairing
channel and describes the procedure for identifying the leading pairing
instability from the matrix structure of the vertex function. In Sec.
IV, we examine the instability of opposite-spin Cooper pairing as
a function of electron filling, considering separately the $d_{xy}$-wave
and $d_{x^{2}-y^{2}}$-wave altermagnetic couplings. Section V presents
a comparison of our results with previous predictions of altermagnetism-induced
FFLO states driven by nearest-neighbor attractive interactions. In
Sec. VI, we discuss the $p$-wave pairing instability for same-spin
electrons, which always occurs at zero center-of-mass momentum, and
contrast it with the opposite-spin pairing case. Finally, Sec. VII
summarizes our findings and outlines possible directions for future
investigation. Appendix A details the spectral broadening procedure
employed in the numerical evaluation of the pair propagators, and
Appendix B summarizes the results obtained upon inclusion of an on-site
attractive interaction.

\section{Model Hamiltonian}

We consider a degenerate electron gas on a square lattice subject
to $d$-wave altermagnetism. In the absence of an external magnetic
field, the system can be modeled by a low-energy single-band Hamiltonian
\citep{Chakraborty2024,Hong2025,Liu2026}, $\mathcal{H}=\mathcal{H}_{0}+\mathcal{H}_{\textrm{int}}$,
of the form,\begin{widetext} 
\begin{gather}
\mathcal{H}_{0}=\sum_{\mathbf{k}}\left[\left(\xi_{\mathbf{k}}+J_{\mathbf{k}}\right)c_{\mathbf{k}\uparrow}^{\dagger}c_{\mathbf{k}\uparrow}+\left(\xi_{\mathbf{k}}-J_{\mathbf{k}}\right)c_{\mathbf{k}\downarrow}^{\dagger}c_{\mathbf{k}\downarrow}\right],\\
\mathcal{H}_{\textrm{int}}=\frac{1}{\mathcal{S}}\sum_{\mathbf{k},\mathbf{k}',\mathbf{q}}V_{\uparrow\downarrow}\left(\mathbf{k},\mathbf{k}'\right)c_{\mathbf{k}+\frac{\mathbf{q}}{2}\uparrow}^{\dagger}c_{\mathbf{-k}+\frac{\mathbf{q}}{2}\downarrow}^{\dagger}c_{\mathbf{-k'}+\frac{\mathbf{q}}{2}\downarrow}c_{\mathbf{k}'+\frac{\mathbf{q}}{2}\uparrow}+\frac{1}{\mathcal{S}}\sum_{\mathbf{k},\mathbf{k}',\mathbf{q};\sigma}V_{\sigma}\left(\mathbf{k},\mathbf{k}'\right)c_{\mathbf{k}+\frac{\mathbf{q}}{2}\sigma}^{\dagger}c_{\mathbf{-k}+\frac{\mathbf{q}}{2}\sigma}^{\dagger}c_{\mathbf{-k'}+\frac{\mathbf{q}}{2}\sigma}c_{\mathbf{k}'+\frac{\mathbf{q}}{2}\sigma},
\end{gather}
\end{widetext}where $c_{\mathbf{k}\sigma}^{\dagger}$ and $c_{\mathbf{k}\sigma}$
represent the creation and annihilation operators, respectively, for
electrons with momentum $\mathbf{k}$ and spin $\sigma=\uparrow,\downarrow$,
and $\mathcal{S}=L^{2}$ is the total number of the lattice sites
for a $L\times L$ square lattice with periodic boundary conditions.

\subsection{Kinetic Hamiltonian}

In the non-interacting kinetic Hamiltonian $\mathcal{H}_{0}$, the
single-particle dispersion is given by 
\begin{equation}
\xi_{\mathbf{k}}\equiv-2t\left(\cos k_{x}+\cos k_{y}\right)-\mu,
\end{equation}
which describes nearest-neighbor hopping with amplitude $t$; the
chemical potential $\mu$ is determined by the electron filling factor
$\nu$. The momentum-dependent spin-splitting $J_{\mathbf{k}}$ originates
from altermagnetism. We focus on the two cases: (i) $d_{xy}$-wave
altermagnetism \citep{Hong2025}, for which
\begin{equation}
J_{\mathbf{k}}=\lambda\sin\left(k_{x}\right)\sin\left(k_{y}\right),
\end{equation}
and (ii) $d_{x^{2}-y^{2}}$-wave altermagnetism \citep{Chakraborty2024},
where
\begin{equation}
J_{\mathbf{k}}=\frac{\lambda}{2}\left[\cos k_{x}-\cos k_{y}\right].
\end{equation}

In real space, the $d_{x^{2}-y^{2}}$-wave altermagnetic order is
realized by introducing spin-dependent nearest-neighbor hopping amplitudes
with opposite signs along the $x$- and $y$-directions \citep{Chakraborty2024}.
In contrast, the $d_{xy}$-wave-wave altermagnetic order arises from
spin-dependent next-nearest-neighbor hopping on a square lattice \citep{Hong2025},
with opposite hopping amplitudes assigned to the two diagonal bond
directions, $(1,1)$ and $(1,-1)$. These distinct bond-dependent
hopping patterns give rise to the characteristic $d_{x^{2}-y^{2}}$-
and $d_{xy}$-wave momentum-space spin splittings, respectively. Therefore,
although both altermagnetic phases originate from bond-dependent spin-dependent
hopping, they differ in both the spatial range of the hopping processes
and the symmetry of the associated bond-order pattern. 

It is important to note that, for both altermagnetic orders, the spin-dependent
hopping term is not invariant under the fourfold lattice rotation
$C_{4}$, which transforms $k_{x}\rightarrow-k_{y}$ and $k_{y}\rightarrow k_{x}$.
Under this rotation, the altermagnetic form factor changes sign, while
the accompanying spin-flip operation reverses the sign of the spin-dependent
hopping term. As a result, the Hamiltonian is invariant under the
combined antiunitary symmetry $C_{4}\mathcal{T}$, even though the
four-fold rotational symmetry $C_{4}$ and time-reversal symmetry
$\mathcal{T}$ are each broken individually. The preservation of the
combined spin-lattice symmetry $C_{4}\mathcal{T}$ is a defining characteristic
of $d$-wave altermagnetic order. 

We further note that, in the dilute-electron limit, the spin-splitting
term reduces to $J_{\mathbf{k}}\propto k_{x}k_{y}$ and $J_{\mathbf{k}}\propto k_{x}^{2}-k_{y}^{2}$
for the $d_{xy}$- and $d_{x^{2}-y^{2}}$-wave altermagnetic orders,
respectively. These two momentum-space form factors are related by
a $45^{\circ}$ rotation, rendering the corresponding low-energy continuum
Hamiltonians equivalent up to a rotation of the coordinates in momentum
space. Consequently, despite their distinct lattice realizations,
the $d_{xy}$- and $d_{x^{2}-y^{2}}$-wave altermagnetic orders are
described by the same free-particle Hamiltonian in the continuum limit
\citep{Cheng2024,Liu2026}.

For notational simplicity, throughout the paper we also define $\xi_{\mathbf{k}\uparrow}=\xi_{\mathbf{k}}+J_{\mathbf{k}}$
and $\xi_{\mathbf{k}\downarrow}=\varepsilon_{\mathbf{k}}-J_{\mathbf{k}}$.

\subsection{Interaction Hamiltonian}

In the interaction Hamiltonian $\mathcal{H}_{\textrm{int}}$, we assume
a short-range interaction potential and adopt the extended attractive
Hubbard $U$-$V$ model, which incorporates both on-site ($U$) and
nearest-neighbor ($V$ and $V_{\sigma}$) pairing interaction terms
\citep{Hong2025}:

\begin{eqnarray}
\mathcal{H}_{\textrm{int}} & = & U\sum_{i}n_{i\uparrow}n_{i\downarrow}+V\sum_{i\delta}n_{i\uparrow}n_{i+\delta,\downarrow}\nonumber \\
 &  & +\frac{V_{\sigma}}{2}\sum_{i\delta}\sum_{\sigma}n_{i\sigma}n_{i+\delta,\sigma},
\end{eqnarray}
where $\delta=\pm\hat{x},\pm\hat{y}$ labels the four nearest neighbors,
and $n_{i\sigma}=c_{i\sigma}^{\dagger}c_{i\sigma}$ denotes the electron
number operator at lattice site $i$. The on-site interaction acts
only between spin-up and spin-down electrons due to the Pauli exclusion
principle. In contrast, the nearest-neighbor interaction can occur
between electrons with either opposite spins or identical spins, characterized
by coupling strengths $V_{\sigma}$ and $V$, respectively, which
are generally distinct.

We emphasize that, in the conventional Hubbard model derived from
the bare Coulomb interaction, both the on-site and nearest-neighbor
interactions are repulsive, with the former typically much larger
than the latter. In contrast, we consider an effective low-energy
interaction Hamiltonian in which the attractive interactions arise
from renormalized pairing mechanisms. Consequently, the effective
interaction parameters are not constrained to follow the hierarchy
of the microscopic Coulomb interactions. For example, spin fluctuations
generated by strong on-site repulsion can mediate an effective nearest-neighbor
attraction \citep{Nakajima1973,Hirsch1985,Scalapino1986}. Similarly,
electron-phonon coupling can induce substantial nearest-neighbor attractive
interactions. Such a scenario has been invoked to explain the anomalously
strong holon-folding spectral feature observed in the quasi-one-dimensional
cuprate Ba$_{2-x}$Sr$_{x}$CuO$_{3+\delta}$, which cannot be reproduced
by the standard Hubbard model without a sizable nearest-neighbor attraction
\citep{Wang2021}. Therefore, an effective regime with $\left|V\right|,\left|V_{\sigma}\right|>U$
is physically well motivated. Unless otherwise specified, we focus
on this parameter regime.

By expressing the interaction Hamiltonian in momentum space using
the standard Fourier transform, we obtain the momentum-dependent nearest-neighbor
interaction strengths \citep{Zhu2023,Chakraborty2024,Hong2025}: 
\begin{eqnarray}
V_{\uparrow\downarrow}\left(\mathbf{k},\mathbf{k}'\right) & = & U+2V\left[\cos\left(k_{x}-k'_{x}\right)+\cos\left(k_{y}-k'_{y}\right)\right],\label{eq:VkkpUnlikeSpin}\\
V_{\sigma}\left(\mathbf{k},\mathbf{k}'\right) & = & V_{\sigma}\left[\cos\left(k_{x}-k'_{x}\right)+\cos\left(k_{y}-k'_{y}\right)\right].\label{eq: VkkpSameSpin}
\end{eqnarray}
The on-site interaction $U$ is purely $s$-wave and momentum-independent,
while the nearest-neighbor interactions ($V$ and $V_{\sigma}$) display
pronounced momentum dependence. These nearest-neighbor interactions
can be decomposed into different symmetry channels - namely, extended
$s$-wave, two $p$-wave, and $d$-wave channels - each of which can
be expressed in a \emph{separable} form \citep{Zhu2023,Hong2025}.
For instance, in the case of opposite spins, we can rewrite,
\begin{equation}
V_{\uparrow\downarrow}\left(\mathbf{k},\mathbf{k}'\right)=\sum_{\eta=\textrm{s,es,\ensuremath{p+},\ensuremath{p-},d}}V_{\eta}f_{\eta}\left(\mathbf{k}\right)f_{\eta}^{*}\left(\mathbf{k}'\right),\label{eq:VkkpDecomposedForm}
\end{equation}
where $V_{\textrm{s}}=U$, $V_{\textrm{es}}=V_{p+}=V_{p-}=V_{\textrm{d}}=V$,
and the form factors of different channels are given by,
\begin{eqnarray}
f_{\textrm{s}}\left(\mathbf{k}\right) & = & 1,\\
f_{\textrm{es}}\left(\mathbf{k}\right) & = & \cos k_{x}+\cos k_{y},\\
f_{p+}\left(\mathbf{k}\right) & = & \sin k_{x}+\sin k_{y},\\
f_{p-}\left(\mathbf{k}\right) & = & \sin k_{x}-\sin k_{y},\\
f_{\textrm{d}}\left(\mathbf{k}\right) & = & \cos k_{x}-\cos k_{y}.
\end{eqnarray}
We can clearly distinguish between the parity-even channel ($\eta=\textrm{s},\textrm{es},\textrm{d}$)
corresponding to spin-singlet pairing, and the parity-odd channel
($\eta=p+,p-$) corresponding to spin-triplet pairing. For the spin-triplet
case, the form factors can alternatively be expressed in the more
familiar chiral forms \citep{Zhu2023,Hong2025}, i.e., $f_{p+ip}(\mathbf{k})=\sin k_{x}+i\sin k_{y}$
and $f_{p-ip}(\mathbf{k})=\sin k_{x}-i\sin k_{y}$. These forms are
not considered in the present study in order to avoid introducing
complex numbers into the numerical calculations.

\begin{figure*}
\begin{centering}
\includegraphics[width=0.8\textwidth]{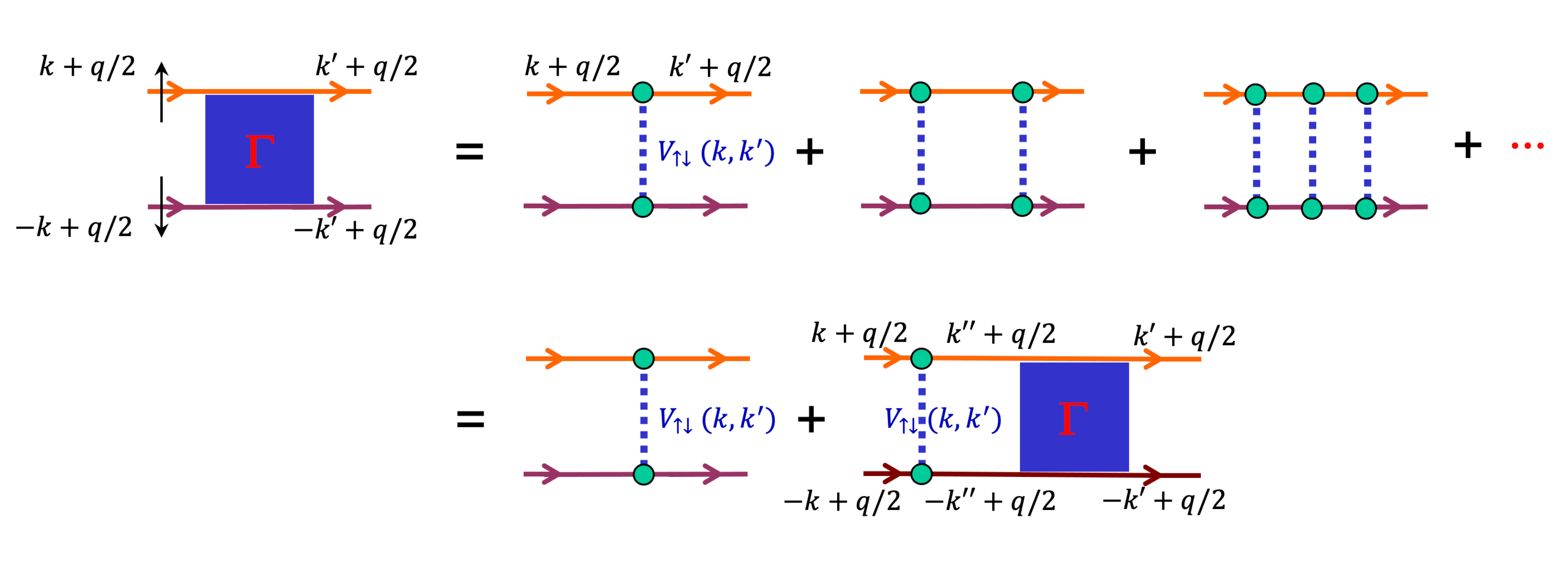}
\par\end{centering}
\caption{\label{fig1} Diagrammatic representation of the two-particle vertex
function $\Gamma(\mathbf{k},\mathbf{k}';\mathbf{Q},\omega)$, within
the standard ladder approximation, for a general non-separable inter-particle
interaction $V_{\uparrow\downarrow}(\mathbf{k},\mathbf{k}')$.}
\end{figure*}

\section{Non-self-consistent $T$-matrix theory and thouless criterion}

We seek to determine the pairing instability of the normal state for
a given set of parameters, including the electron filling $\nu$,
the interaction strengths $U$, $V$ and $V_{\sigma}$, and the altermagnetic
coupling $\lambda$. This can be conveniently analyzed using the Thouless
criterion applied to the inverse two-particle vertex function $\Gamma^{-1}(\mathbf{Q},\omega=0)$.
In the normal state, the inverse vertex function is negative, while
in the superconducting phase it becomes positive. The superconducting
phase transition occurs when it crosses zero \citep{Thouless1960,Liu2006},
\begin{equation}
\max_{\{\mathbf{Q}\}}\Gamma^{-1}\left(\mathbf{Q},\omega=0\right)=0.\label{eq:ThoulessCriterion}
\end{equation}
If the maximum of the inverse vertex function takes place at a nonzero
momentum $\mathbf{Q}\neq0$, the normal state becomes unstable toward
the formation of an inhomogeneous FFLO superconducting state. 

We note that the vertex function is not necessarily a scalar quantity.
When multiple pairing channels are present, it generally takes a matrix
form. In such cases, one should consider the \emph{largest} eigenvalue
$\gamma_{1}(\mathbf{Q})$ of the inverse vertex function matrix at
zero frequency. Accordingly, in the Thouless criterion, $\Gamma^{-1}(\mathbf{Q},\omega=0)$
is replaced by $\gamma_{1}(\mathbf{Q})$. In the following, we employ
the standard non-self-consistent many-body $T$-matrix approach to
evaluate the two-particle vertex function \citep{MahanManyParticlePhysics}.

\subsection{Pairing between electrons with opposite spins}

Restricting our attention first to pairing between electrons with
\emph{opposite} spins, we formally derive the equation for the vertex
function $\Gamma(\mathbf{k},\mathbf{k}';\mathbf{Q},\omega)$ in the
Cooperon channel within the ladder approximation. By summing the diagrams
shown in Fig. \ref{fig1}, we obtain,\begin{widetext}
\begin{equation}
\Gamma\left(\mathbf{k},\mathbf{k}';\mathbf{Q},\omega\right)=V_{\uparrow\downarrow}\left(\mathbf{k},\mathbf{k}'\right)-\sum_{k''}V_{\uparrow\downarrow}\left(\mathbf{k},\mathbf{k}''\right)G_{0\uparrow}\left(k''+\frac{q}{2}\right)G_{0\downarrow}\left(-k''+\frac{q}{2}\right)\Gamma\left(\mathbf{k}'',\mathbf{k}';\mathbf{Q},\omega\right),
\end{equation}
where $G_{0\uparrow}$ and $G_{0\downarrow}$ denote the non-interacting
single-particle Green's functions, which depend on the four momenta
$k''\equiv\{\mathbf{k}'',i\omega_{m}\}$ and $q\equiv\{\mathbf{Q},i\nu_{n}\rightarrow\omega^{+}\equiv\omega+i0^{+}\}$.
The summation $\sum_{k''}\equiv(k_{B}T/\mathcal{S})\sum_{i\omega_{m}}\sum_{\mathbf{k}''}$
runs over both the spatial momentum and the (fermionic) Matsubara
frequency $\omega_{m}=(2m+1)\pi k_{B}T$, with integer $m\in\mathbb{Z}$.
The summation over $\omega_{m}$ can be performed straightforwardly
\citep{MahanManyParticlePhysics}, yielding the expression,
\begin{equation}
\Gamma\left(\mathbf{k},\mathbf{k}';\mathbf{Q},\omega\right)=V_{\uparrow\downarrow}\left(\mathbf{k},\mathbf{k}'\right)-\frac{1}{\mathcal{S}}\sum_{\mathbf{k}''}V_{\uparrow\downarrow}\left(\mathbf{k},\mathbf{k}''\right)\frac{\left[n_{F}\left(\xi_{\frac{\mathbf{Q}}{2}+\mathbf{k''}\uparrow}\right)+n_{F}\left(\xi_{\frac{\mathbf{Q}}{2}-\mathbf{k''}\downarrow}\right)-1\right]}{\omega^{+}-\left(\xi_{\frac{\mathbf{Q}}{2}+\mathbf{k''}\uparrow}+\xi_{\frac{\mathbf{Q}}{2}-\mathbf{k}''\downarrow}\right)}\Gamma\left(\mathbf{k}'',\mathbf{k}';\mathbf{Q},\omega\right),\label{eq:VertexEquation}
\end{equation}
where $n_{F}(x)\equiv1/(e^{x/k_{B}T}+1)$ is the Fermi-Dirac distribution
function. In general, this equation is challenging to solve. However,
by decomposing the interaction potential as in Eq. (\ref{eq:VkkpDecomposedForm})
into a total of $N$ channels (with $N=5$ in the present case, corresponding
to $\eta=\textrm{s,es,\ensuremath{p+},\ensuremath{p-},d}$, or equivalently
$\eta=1,...,5$),
\begin{equation}
V_{\uparrow\downarrow}\left(\mathbf{k},\mathbf{k}'\right)=\left[\begin{array}{ccccc}
f_{1}\left(\mathbf{k}\right), & \cdots, & f_{\eta}\left(\mathbf{k}\right), & \cdots, & f_{N}(\mathbf{k})\end{array}\right]\left[\begin{array}{ccccc}
V_{1}\\
 & \ddots\\
 &  & V_{\eta}\\
 &  &  & \ddots\\
 &  &  &  & V_{N}
\end{array}\right]\left[\begin{array}{c}
f_{1}^{*}\left(\mathbf{k}'\right)\\
\vdots\\
f_{\eta}^{*}\left(\mathbf{k}'\right)\\
\vdots\\
f_{N}^{*}\left(\mathbf{k}'\right)
\end{array}\right],
\end{equation}
we may rewrite the vertex function in the form of,
\[
\Gamma\left(\mathbf{k},\mathbf{k}';\mathbf{Q},\omega\right)=\left[\begin{array}{ccccc}
f_{1}\left(\mathbf{k}\right), & \cdots, & f_{\eta}\left(\mathbf{k}\right), & \cdots, & f_{N}(\mathbf{k})\end{array}\right]\boldsymbol{\Gamma}\left(\mathbf{Q},\omega\right)\left[\begin{array}{c}
f_{1}^{*}\left(\mathbf{k}'\right)\\
\vdots\\
f_{\eta}^{*}\left(\mathbf{k}'\right)\\
\vdots\\
f_{N}^{*}\left(\mathbf{k}'\right)
\end{array}\right],
\]
where $\boldsymbol{\Gamma}(\mathbf{Q},\omega)$ is a $N\times$$N$
matrix. Thus, by introducing a $N\times$$N$ pair propagator matrix
$\boldsymbol{\chi}(\mathbf{Q},\omega)$, whose matrix elements are
given by 
\begin{equation}
\chi_{\eta\eta'}\left(\mathbf{Q},\omega\right)\equiv\frac{1}{\mathcal{S}}\sum_{\mathbf{k}''}f_{\eta}^{*}\left(\mathbf{k}''\right)\frac{\left[n_{F}\left(\xi_{\frac{\mathbf{Q}}{2}+\mathbf{k''}\uparrow}\right)+n_{F}\left(\xi_{\frac{\mathbf{Q}}{2}-\mathbf{k''}\downarrow}\right)-1\right]}{\omega^{+}-\left(\xi_{\frac{\mathbf{Q}}{2}+\mathbf{k''}\uparrow}+\xi_{\frac{\mathbf{Q}}{2}-\mathbf{k}''\downarrow}\right)}f_{\eta'}\left(\mathbf{k}''\right),\label{eq:PairPropagator}
\end{equation}
we can recast the vertex equation, Eq. (\ref{eq:VertexEquation}),
in matrix form for $\boldsymbol{\Gamma}(\mathbf{Q},\omega)$,
\begin{equation}
\boldsymbol{\Gamma}\left(\mathbf{Q},\omega\right)=\mathbf{V}-\mathbf{V}\boldsymbol{\chi}\left(\mathbf{Q},\omega\right)\mathbf{\boldsymbol{\Gamma}}\left(\mathbf{Q},\omega\right),
\end{equation}
where the interaction matrix $\mathbf{V}\equiv\textrm{diag}\{V_{1},\cdots,V_{\eta},\cdots,V_{N}\}$
is diagonal. It is readily seen that we can formally solve for the
matrix of the inverse vertex function,
\begin{equation}
\boldsymbol{\Gamma}^{-1}\left(\mathbf{Q},\omega\right)=\mathbf{V}^{-1}+\boldsymbol{\chi}\left(\mathbf{Q},\omega\right)=\left[\begin{array}{ccccc}
V_{1}^{-1}+\chi_{11}\left(\mathbf{Q},\omega\right) & \cdots & \chi_{1\eta}\left(\mathbf{Q},\omega\right) & \cdots & \chi_{1N}\left(\mathbf{Q},\omega\right)\\
\vdots & \ddots & \cdots & \cdots & \vdots\\
\chi_{\eta1}\left(\mathbf{Q},\omega\right) & \cdots & V_{\eta}^{-1}+\chi_{\eta\eta}\left(\mathbf{Q},\omega\right) & \cdots & \chi_{\eta N}\left(\mathbf{Q},\omega\right)\\
\vdots & \cdots & \vdots & \ddots & \vdots\\
\chi_{N1}\left(\mathbf{Q},\omega\right) & \cdots & \chi_{N\eta}\left(\mathbf{Q},\omega\right) & \cdots & V_{N}^{-1}+\chi_{NN}\left(\mathbf{Q},\omega\right)
\end{array}\right].\label{eq:VertexFunction5}
\end{equation}
\end{widetext}If the off-diagonal elements $\chi_{\eta\neq\eta'}(\mathbf{Q},\omega=0)$
are negligible, the different channels effectively decouple, and it
suffices to select the channel with the largest diagonal element.
In general, however, for each center-of-mass pairing momentum $\mathbf{Q}$,
one must diagonalize the full matrix $\boldsymbol{\Gamma}^{-1}(\mathbf{Q},\omega=0)$
to determine its largest eigenvalue, i.e., $\gamma_{1}(\mathbf{Q})$,
which identifies the \emph{leading} pairing instability in the Thouless
criterion that now takes the form,

\begin{equation}
\max_{\{\mathbf{Q}\}}\gamma_{1}\left(\mathbf{Q}\right)=0.\label{eq:ThoulessCriterionNow}
\end{equation}
Therefore, the superconducting instability sets in once $\gamma_{1}(\mathbf{Q})>0$
at some pairing momenta $\mathbf{Q}$. The associated eigenvector
$[M_{1},\cdots,M_{\eta},\cdots,M_{N}]{}^{T}$ then specifies the structure
of the resulting superconducting order parameter. The remaining eigenvalues
of $\boldsymbol{\Gamma}^{-1}(\mathbf{Q},\omega=0)$ are denoted by
$\gamma_{n}(\mathbf{Q})$, ordered in decreasing magnitude for $n=2,\cdots,N$.
The second-largest eigenvalue, $\gamma_{2}(\mathbf{Q})$, and the
corresponding eigenvector provide insight into the nature of the potential
competing superconducting instability.

An alternative approach to determining the superconducting instability
is to solve the linearized gap equation $\mathbf{K}\Delta=\Delta$,
which can be formulated as an eigenvalue problem for the pairing kernel
$\mathbf{K}=-\mathbf{V}\mathbf{\boldsymbol{\chi}}(\mathbf{Q},\omega=0)$.
The superconducting transition is reached when the largest eigenvalue
of the kernel becomes unity, while the corresponding eigenvector determines
the momentum dependence and symmetry of the superconducting gap. By
contrast, the present work employs the Thouless criterion, which characterizes
the pairing instability through the divergence of the two-particle
(pairing) vertex function $\mathbf{\Gamma}(\mathbf{Q},\omega=0)$,
or equivalently the vanishing of the largest eigenvalue of the inverse
two-particle vertex function. Although the two approaches emphasize
different physical perspectives - the former treating the pairing
instability as an eigenvalue problem of the gap equation and the latter
interpreting it as a softening of the two-particle vertex - they are
mathematically equivalent. Indeed, since the inverse two-particle
vertex function can be written as $\boldsymbol{\Gamma}^{-1}(\mathbf{Q},\omega=0)=\mathbf{V}^{-1}(\mathbf{I}-\mathbf{K})$,
the condition that the largest eigenvalue of the pairing kernel $\mathbf{K}$
reaches unity is precisely equivalent to the vanishing of the largest
eigenvalue of $\boldsymbol{\Gamma}^{-1}$. Consequently, both approaches
predict the same superconducting phase boundary and identify the same
pairing symmetry through the corresponding critical eigenvector. The
Thouless criterion, however, naturally expresses the instability in
terms of the normal-state response function and is therefore particularly
well suited for analyzing finite-momentum pairing instabilities discussed
in this work.

\subsection{Two-electron limit}

At this point, it is instructive to consider the two-electron limit,
where the two Fermi surfaces are absent. This corresponds to omitting
the two Fermi--Dirac distribution functions in the pair propagators
of Eq. (\ref{eq:PairPropagator}). By absorbing the chemical potential
into the frequency and defining the energy $E=\omega+2\mu$, we introduce
\begin{equation}
L_{\eta\eta'}\equiv-\chi_{\eta\eta'}=\frac{1}{\mathcal{S}}\sum_{\mathbf{k}}\frac{f_{\eta}^{*}\left(\mathbf{k}\right)f_{\eta'}\left(\mathbf{k}\right)}{E-\left(\varepsilon_{\frac{\mathbf{Q}}{2}+\mathbf{k}\uparrow}+\varepsilon_{\frac{\mathbf{Q}}{2}-\mathbf{k}\downarrow}\right)},
\end{equation}
where $\varepsilon_{\mathbf{k}\sigma}=\xi_{\mathbf{k}\sigma}+\mu$.
The inverse vertex matrix can then be written as, $\boldsymbol{\Gamma}_{\eta\eta'}^{-1}(\mathbf{Q},E)=-L_{\eta\eta'}(E)+V_{\eta}^{-1}\delta_{\eta\eta'}$.
Physically, $\boldsymbol{\Gamma}_{\eta\eta'}(\mathbf{Q},\omega)$
may be interpreted as the Green's function of a Cooper pair, whose
poles correspond to quasi-bound states with energies measured relative
to the Fermi surfaces. Therefore, in the two-electron limit, the zeros
of $\boldsymbol{\Gamma}_{\eta\eta'}^{-1}(\mathbf{Q},E)$, determined
by the condition,
\begin{equation}
\det\left[L_{\eta\eta'}\left(E\right)-\frac{\delta_{\eta\eta'}}{V_{\eta}}\right]=0,\label{eq:det}
\end{equation}
give rise to the energies of tightly bound Cooper pairs. This reproduces
the central equation previously derived for the two-electron bound-state
problem \citep{Hu2026}.

Returning to the many-body case, condensation of these bound pairs
occurs when the pair energy $E$ approaches the pair chemical potential
$2\mu$, i.e., when $\omega=0$, provided that residual interactions
between bound pairs are neglected and $\boldsymbol{\Gamma}_{\eta\eta'}(\mathbf{Q},\omega)$
is regarded as the Green's function of noninteracting composite bosons
\citep{MahanManyParticlePhysics}. In this way, the Thouless criterion
naturally emerges, following the standard picture of Bose--Einstein
condensation in an ideal Bose gas.

\subsection{Pairing with parallel spins}

We now consider pairing between electrons with \emph{identical} spins.
In this situation, the symmetric spin-triplet structure in spin space
requires the real-space wave function to have odd parity. Consequently,
only the two $p$-wave channels in the interaction potential $V_{\sigma}\left(\mathbf{k},\mathbf{k}'\right)$
contribute. Following an analogous many-body $T$-matrix procedure,
we obtain the $2\times2$ inverse vertex matrix,
\begin{equation}
\boldsymbol{\Gamma}_{\sigma}^{-1}\left(\mathbf{Q},\omega\right)=\left[\begin{array}{cc}
V_{\sigma}^{-1}+\chi_{11}^{\sigma} & \chi_{12}^{\sigma}\\
\chi_{21}^{\sigma} & V_{\sigma}^{-1}+\chi_{22}^{\sigma}
\end{array}\right],\label{eq:VertexFunction2}
\end{equation}
where the indices $\eta,\eta'=1,2$ now label the $p+$ and $p-$
channels, respectively. The pair propagator $\chi_{\eta\eta'}^{\sigma}$
can again be evaluated using Eq. (\ref{eq:PairPropagator}), but with
the identical single-particle dispersion for the two electrons, namely
either $\xi_{\mathbf{Q}/2\pm\mathbf{k}'',\uparrow}$ or $\xi_{\mathbf{Q}/2\pm\mathbf{k}'',\downarrow}$. 

\section{Leading pairing instability}

The numerical procedure for identifying the leading pairing instability
from the vertex matrices, Eq. (\ref{eq:VertexFunction5}) and Eq.
(\ref{eq:VertexFunction2}), is straightforward. The primary numerical
difficulty arises when evaluating the pair propagator in Eq. (\ref{eq:PairPropagator}).
Specifically, when the denominator approaches zero, the integrand
develops a sharp peak at low temperature. To regularize this feature,
a small spectral broadening factor $\eta$ is introduced. The final
results are then extrapolated to the limit $\eta=0^{+}$ using a cubic
fit. As examined in detail in Appendix A, this strategy is efficient
and introduces only negligible numerical errors.

Unless otherwise specified, we perform our calculations at an effectively
zero temperature, $T=0.01t$, in order to smooth the sharp Fermi surfaces.
We assume a negligible on-site attraction, $U=-0.01t$, along with
nearest-neighbor attractions $V=-1.5t$ and $V_{\sigma}=-1.5t$. The
impact of varying the on-site attraction $U$ on the pairing instability
is discussed in Appendix B.

\begin{figure}
\begin{centering}
\includegraphics[width=0.5\textwidth]{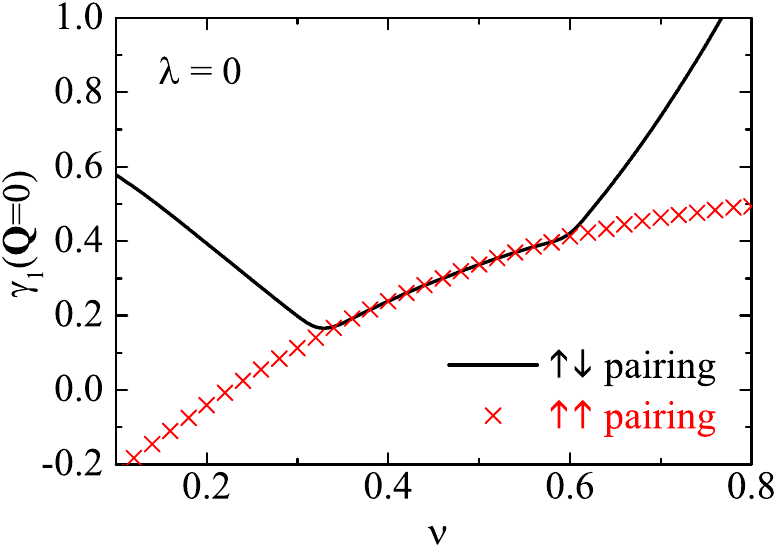}
\par\end{centering}
\caption{\label{fig2} The largest eigenvalue of the inverse vertex function
matrix, $\gamma_{1}(\mathbf{Q}=0)$, is plotted as a function of the
lattice filling factor $\nu$, in the case without altermagnetism
(i.e., $\lambda=0$). The curves represent the results for pairing
between electrons with unlike spins, while the red crosses indicate
the triplet pairing results for spin-up electrons. Here and in the
following figures, we adopt the interaction parameters $U=-0.01t$
and $V=V_{\sigma}=-1.5t$, and temperature $T=0.01t$, unless otherwise
stated.}
\end{figure}

\subsection{The case without altermagnetism}

We begin by analyzing the pairing instability in the absence of altermagnetism,
where it consistently occurs at $\mathbf{Q}=0$. In Fig. \ref{fig2},
we present the leading eigenvalue, $\gamma_{1}(\mathbf{Q}=0)$, of
the inverse vertex matrix $\mathbf{\Gamma}^{-1}$ for antiparallel
spins (black solid line) and $\mathbf{\Gamma}_{\sigma}^{-1}$ for
parallel spins (red crosses), plotted as a function of the electron
filling factor $\nu$. 

For antiparallel-spin pairing, the leading eigenvalue initially decreases
with increasing $\nu$, then gradually rises around $\nu\simeq0.3$,
and eventually grows more steeply once $\nu\apprge0.6$. This non-monotonic
behavior reflects changes in the dominant pairing channel \citep{Romer2015}.
At low electron filling, the leading instability corresponds to an
extended $s$-wave state. In the intermediate regime, $0.3\lesssim\nu\lesssim0.6$,
it transitions to a $p$-wave state. Finally, for $\nu\apprge0.6$,
the pairing symmetry switches to $d$-wave. This interpretation is
supported by the behavior of the leading eigenvalue for triplet $p$-wave
pairing among spin-up electrons, which coincides in the intermediate
filling range $0.3\lesssim\nu\lesssim0.6$, owing to the identical
nearest-neighbor interaction strengths $V=V_{\sigma}$.

Notably, in the absence of altermagnetism, the leading eigenvalue
for antiparallel-spin pairing is always at least as large as that
for parallel-spin pairing. This indicates that parallel-spin pairing
does not constitute the dominant instability for sufficiently small
altermagnetic couplings when $\left|V\right|\geq\left|V_{\sigma}\right|$.
Accordingly, in this section and the next, we focus on antiparallel-spin
pairing. In the following, we separately discuss the two cases of
the $d_{xy}$- and $d_{x^{2}-y^{2}}$-wave altermagnetism.

\begin{figure}
\begin{centering}
\includegraphics[width=0.5\textwidth]{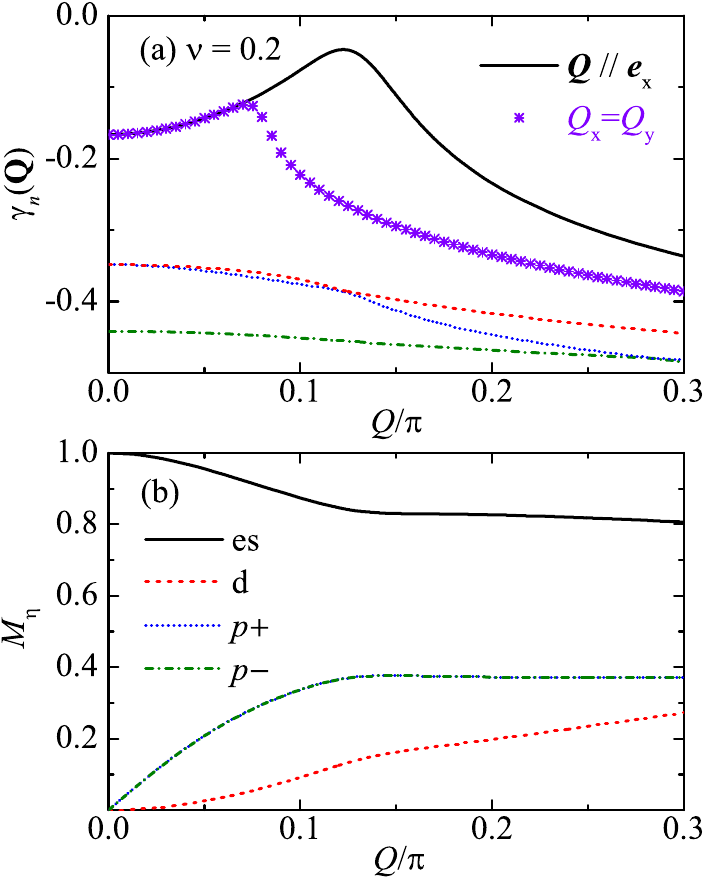}
\par\end{centering}
\caption{\label{fig3} (a) The curves represent the four eigenvalues of the
inverse vertex function matrix, $\gamma_{n}(\mathbf{Q})$, at the
filling factor $\nu=0.2$, as a function of the center-of-mass momentum
$\mathbf{Q}$ when it lies along the $x$-axis. The crosses show $\gamma_{1}(\mathbf{Q})$
for $\mathbf{Q}$ directed along the diagonal ($Q_{x}=Q_{y}$). In
both orientations, introducing a $d_{xy}$-wave altermagnetic coupling
with strength $\lambda=0.5t$ causes $\gamma_{1}(\mathbf{Q})$ to
peak at a nonzero $\mathbf{Q}$. (b) The eigenvectors associated with
$\gamma_{1}(\mathbf{Q})$ are shown for $\mathbf{Q}$ along the $x$-axis. }
\end{figure}

\subsection{$d_{xy}$-wave altermagnetism}

In Fig. \ref{fig3}(a), we report the eigenvalues of the inverse vertex
matrix $\gamma_{n}(\mathbf{Q})$ for a $d_{xy}$-wave altermagnetic
coupling $\lambda=0.5t$ at low electron filling $\nu=0.2$, plotted
as a function of the center-of-mass momentum $Q=\left|\mathbf{Q}\right|$
with $\mathbf{Q}$ oriented along the $x$-direction. Only four eigenvalues
are visible, since the fifth one - associated with the $s$-wave channel
- lies far below the others due to the very weak on-site attraction
$U=-0.01t$. The leading eigenvalue $\gamma_{1}(\mathbf{Q})$ exhibits
a pronounced maximum at a finite momentum $Q\simeq0.12\pi$. For other
directions of $\mathbf{Q}$, $\gamma_{1}(\mathbf{Q})$ likewise develops
a local maximum at nonzero $Q$. The case where $\mathbf{Q}$ points
along the diagonal direction (i.e., $Q_{x}=Q_{y}$) is explicitly
indicated in the figure by purple crosses. Nevertheless, in the presence
of $d_{xy}$-wave altermagnetism, the global maximum of the leading
eigenvalue consistently occurs when $\mathbf{Q}$ is aligned with
the $x$-direction. 

In general, the leading eigenvalue is associated with an eigenvector
whose components $M_{\ensuremath{\eta}}$ characterize the relative
weight of different pairing channels in the dominant pairing instability.
Figure \ref{fig3}(b) shows these components as a function of $Q$.
At the peak momentum $Q\simeq0.12\pi$, the extended $s$-wave channel
clearly dominates, accompanied by two weaker degenerate $p$-wave
contributions and only a negligible component from the $d$-wave channel.

\begin{figure}
\begin{centering}
\includegraphics[width=0.5\textwidth]{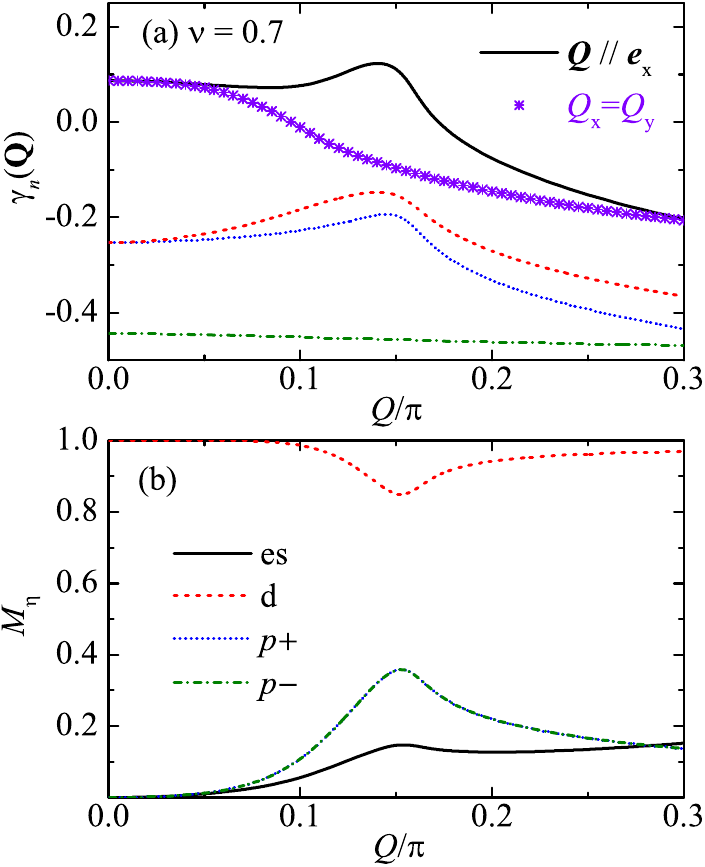}
\par\end{centering}
\caption{\label{fig4} The results are identical to that presented in Fig.
\ref{fig3}, with the exception that they are calculated at a larger
lattice filling factor of $\nu=0.7$.}
\end{figure}

At higher electron fillings, illustrated in Fig. \ref{fig4} for the
representative case $\nu=0.7$, the leading eigenvalue of the inverse
vertex matrix $\gamma_{1}(\mathbf{Q})$ displays behavior qualitatively
similar to that at low filling, but with two important differences.
On the one hand, although the global maximum still happens at nonzero
momentum when $\mathbf{Q}$ directs along the $x$-direction, the
point $Q=0$ now becomes a local maximum. This feature is especially
evident when $Q$ lies along the diagonal: in that case, the finite-momentum
peak disappears entirely (see purple crosses). On the other hand,
as the electron filling increases, the structure of the eigenvector
associated with the leading eigenvalue changes dramatically. At the
peak momentum $Q\simeq0.15\pi$, the pairing instability is dominated
by a $d$-wave component, followed by two degenerate $p$-wave channels,
while the extended $s$-wave contribution becomes negligible. This
evolution of the dominant pairing channel with increasing filling
is consistent with the trend observed in the absence of altermagnetism,
as previously shown in Fig. \ref{fig2}.

\begin{figure}
\begin{centering}
\includegraphics[width=0.5\textwidth]{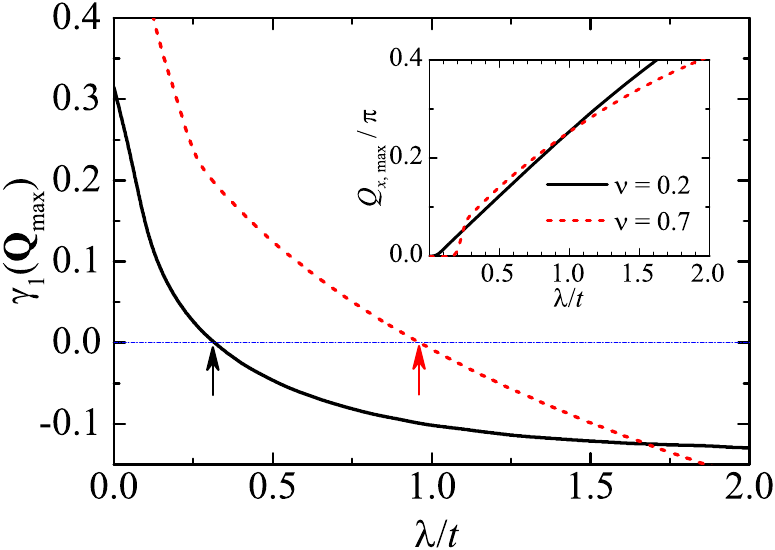}
\par\end{centering}
\caption{\label{fig5} The peak value of the dominant eigenvalue of the inverse
vertex function matrix, $\gamma_{1}(\mathbf{Q}_{\textrm{max}})$,
is shown as a function of the $d_{xy}$-wave altermagnetic coupling
strength $\lambda$ for two lattice filling factors: $\nu=0.2$ (black
solid line) and $\nu=0.7$ (red dashed line). The arrows highlight
the critical altermagnetic coupling at which $\gamma_{1}(\mathbf{Q}_{\textrm{max}})$
vanishes. The inset shows the magnitude of the $\mathbf{Q}$ vector
at which this peak occurs.}
\end{figure}

We have systematically repeated the numerical calculations for various
values of altermagnetic couplings $\lambda$ and electron fillings
$\nu$, in order to determine the global maximum of the leading eigenvalue
of the inverse vertex matrix $\gamma_{1}(\mathbf{Q}_{\textrm{max}})$,
the corresponding peak momentum $\mathbf{Q}_{\textrm{max}}$, and
the weights of the different pairing channels, $W_{\eta}=\left|M_{\eta}\right|^{2}$,
obtained from the associated eigenvector at $\mathbf{Q}_{\textrm{max}}$.
In Fig. \ref{fig5}, we plot the global maximum as a function of $\lambda$
for two representative filling factors: $\nu=0.2$ (black solid line)
and $\nu=0.7$ (red dashed line), together with the corresponding
peak momentum shown in the inset. 

As $\lambda$ increases, $\gamma_{1}(\mathbf{Q}_{\textrm{max}})$
generally decreases. This behavior can be understood from the fact
that altermagnetism acts as a momentum-dependent effective magnetic
field, which tends to suppress Cooper pairing. As a consequence, with
increasing $\lambda$, the maximum may eventually reach zero at a
critical altermagnetic coupling $\lambda_{c}$. According to the Thouless
criterion Eq. (\ref{eq:ThoulessCriterion}), this critical point marks
the boundary between the superconducting phase and the normal state.
In the main panel, the critical couplings $\lambda_{c}$ for $\nu=0.2$
and $\nu=0.7$ are indicated by black and red arrows, respectively.
The inset further shows that the superconducting phase transition
is always associated with a nonzero peak momentum, $\mathbf{Q}_{\textrm{max}}\neq0$.
This demonstrates that the emergent superconducting phase corresponds
to a finite-momentum FFLO state.

\begin{figure}
\begin{centering}
\includegraphics[width=0.5\textwidth]{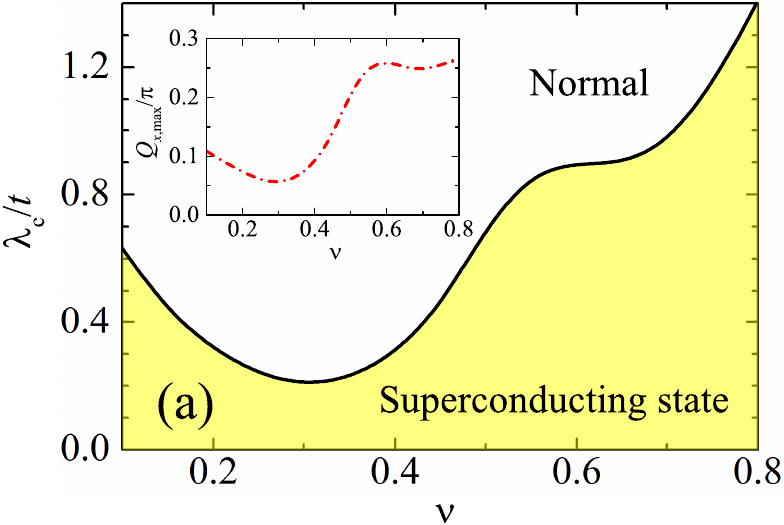}
\par\end{centering}
\begin{centering}
\includegraphics[width=0.5\textwidth]{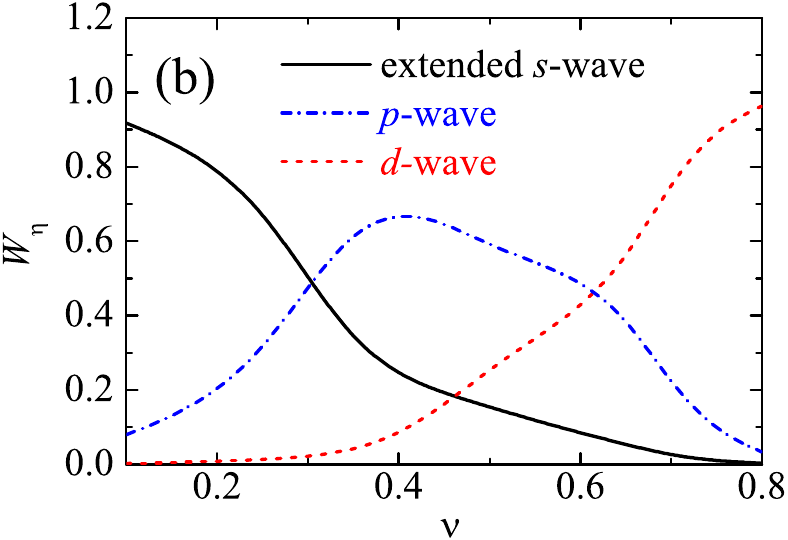}
\par\end{centering}
\caption{\label{fig6} (a) Phase diagram in the presence of $d_{xy}$-wave
altermagnetism: the critical altermagnetic coupling strength $\lambda_{c}$
is plotted as a function of the lattice filling factor $\nu$. The
inset shows the center-of-mass momentum $\mathbf{Q}=Q\mathbf{e}_{x}$
at the superconducting phase boundary. (b) Contribution of various
partial-wave components to the leading pairing channel, $W_{\eta}=\left|M_{\eta}\right|^{2}$,
is shown as a function of $\nu$ at the superconducting transition.}
\end{figure}

\begin{figure}
\begin{centering}
\includegraphics[width=0.5\textwidth]{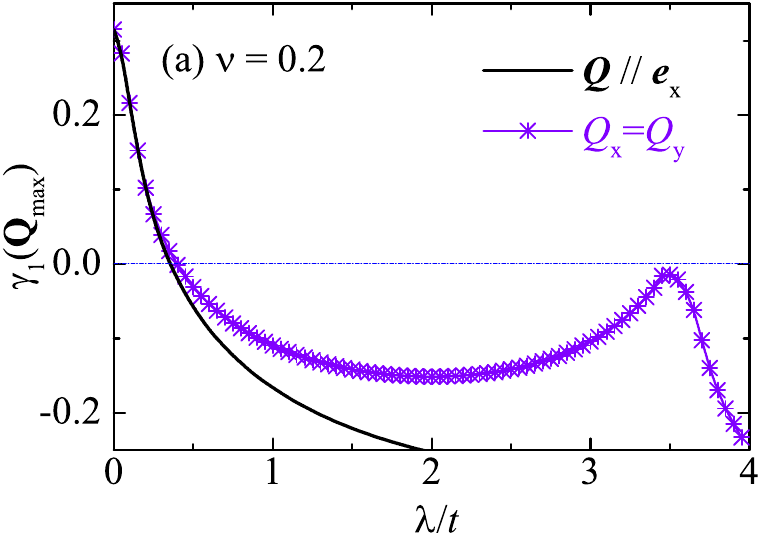}
\par\end{centering}
\begin{centering}
\includegraphics[width=0.5\textwidth]{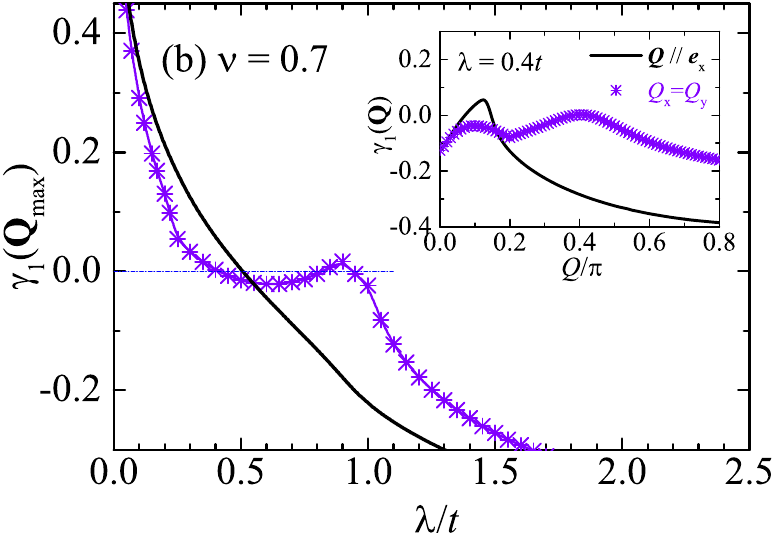}
\par\end{centering}
\caption{\label{fig7} The peak value of the leading eigenvalue of the inverse
vertex function matrix, $\gamma_{1}(\mathbf{Q}_{\textrm{max}})$,
is plotted as a function of the $d_{x^{2}-y^{2}}$-wave altermagnetic
coupling strength $\lambda$ for two lattice filling factors: $\nu=0.2$
(top panel) and $\nu=0.7$ (bottom panel). In each panel, the black
curve corresponds to the center-of-mass momentum $\mathbf{Q}$ aligned
along the $x$-axis, while the violet crosses represent $\mathbf{Q}$
oriented along the $x$-axis or the diagonal ($Q_{x}=Q_{y}$). The
inset in (b) shows $\gamma_{1}(\mathbf{Q})$ for different orientations
of Q, emphasizing that at small $d_{x^{2}-y^{2}}$-wave altermagnetic
couplings $\gamma_{1}(\mathbf{Q}_{\textrm{max}})$ occurs when $\mathbf{Q}$
is directed along the $x$-axis.}
\end{figure}

In Fig. \ref{fig6}(a), we present the critical altermagnetic coupling
$\lambda_{c}$ as a function of electron filling $\nu$, thereby establishing
the corresponding phase diagram. The associated peak momentum $Q_{\textrm{max}}$
is displayed in the inset. Both $\lambda_{c}$ and $Q_{\textrm{max}}$
show a similar non-monotonic dependence on $\nu$, closely resembling
the behavior observed previously in Fig. \ref{fig2} in the absence
of altermagnetism. For instance, as $\nu$ increases, the initially
decreasing trend of $\lambda_{c}$ reverses around $\nu\simeq0.3$.
Remarkably, this is precisely the filling at which - without altermagnetism
- the dominant pairing channel switches from extended $s$-wave to
$p$-wave symmetry. Furthermore, $\lambda_{c}$ exhibits another change
in slope and develops a plateau near $\nu\simeq0.6$, where Fig. \ref{fig2}
indicates a transition of the dominant pairing channel from $p$-wave
to $d$-wave character. 

Motivated by this correspondence, it is natural to relate the non-monotonic
$\nu$-dependence of $\lambda_{c}$ to changes in the nature of the
leading pairing instability. To examine this connection, Fig. \ref{fig6}(b)
shows the weights of the various pairing channels evaluated precisely
at the superfluid-normal phase boundary. Indeed, with increasing $\nu$
the dominant pairing symmetry evolves from extended $s$-wave to $p$-wave
at $\nu\simeq0.3$, and subsequently to $d$-wave at $\nu\simeq0.6$.
We emphasize, however, that in the presence of altermagnetism the
sub-leading pairing channels remain substantial, particularly near
the crossover points at $\nu\simeq0.3$ and $\nu\simeq0.6$. Consequently,
once the superconducting phase emerges, the pairing order parameter
is generally multi-component in nature, incorporating contributions
from both spin-singlet and spin-triplet channels.

\begin{figure}
\begin{centering}
\includegraphics[width=0.5\textwidth]{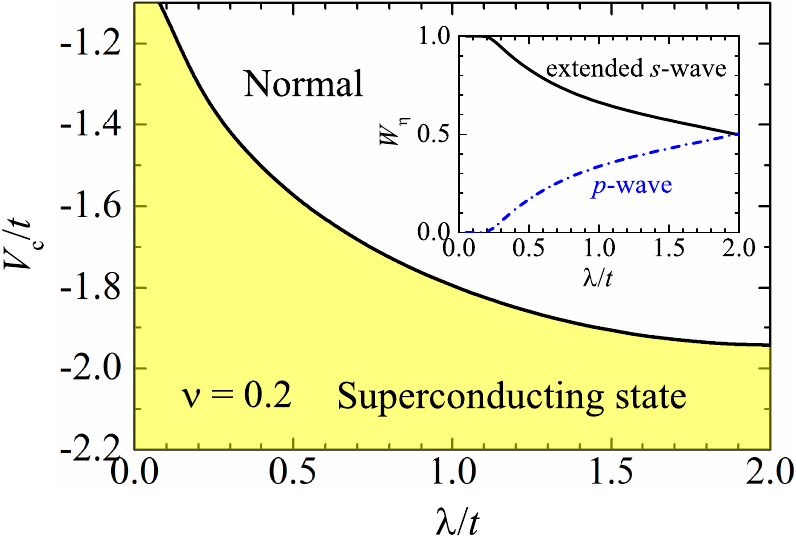}
\par\end{centering}
\caption{\label{fig8} Phase diagram with $d_{x^{2}-y^{2}}$-wave altermagnetism:
the critical nearest-neighbor interaction strength $V_{c}$ is shown
as a function of the altermagnetic coupling strength $\lambda$, at
the lattice filling factor $\nu=0.2$. The inset displays the contributions
of different partial-wave components to the dominant pairing channel,
$W_{\eta}=\left|M_{\eta}\right|^{2}$, at the superconducting transition.
Here, we take the on-site attraction $U=-0.01t$.}
\end{figure}

\subsection{$d_{x^{2}-y^{2}}$-wave altermagnetism}

We now proceed to analyze the case of $d_{x^{2}-y^{2}}$-wave altermagnetism.
Fig. \ref{fig7}(a) and Fig. \ref{fig7}(b) report $\gamma_{1}(\mathbf{Q}_{\textrm{max}})$
as a function of the altermagnetic coupling $\lambda$, at two electron
fillings $\nu=0.2$ and $\nu=0.7$, respectively. As in the previous
analysis, we consider two orientations of the center-of-mass momentum
$\mathbf{Q}$: aligned either along the $x$-axis (black curves) or
along the diagonal direction (purple crosses). 

For low electron filling ($\nu=0.2$), we observe that $\gamma_{1}(\mathbf{Q}_{\textrm{max}})$
occurs when $\mathbf{Q}$ points along the diagonal direction, in
clear contrast to the behavior found for $d_{xy}$-wave altermagnetism.
At higher filling ($\nu=0.7$), the same tendency persists when the
altermagnetic coupling is sufficiently strong. However, for weaker
couplings, specifically $\lambda\leq0.55t$, the maximum of $\gamma_{1}(\mathbf{Q})$
instead appears $\mathbf{Q}$ lies along the $x$-axis, resembling
the behavior characteristic of the $d_{xy}$-wave case. This re-orientation
of the optimal momentum $\mathbf{Q}$ is most clearly illustrated
by examining the $Q$-dependence of $\gamma_{1}(\mathbf{Q})$ at a
relatively small coupling $\lambda=0.4t$, as shown in the inset of
Fig. \ref{fig7}(b). In this regime, we typically find a single peak
when $\mathbf{Q}$ is directed along the $x$-axis, whereas a double-peak
structure emerges when $\mathbf{Q}$ lies along the diagonal. In general,
the single peak in the former configuration competes with the outer
peak of the doublet in the latter, and becomes energetically favorable
at weaker altermagnetic coupling.

In contrast to the $d_{xy}$-wave altermagnetism, a notable feature
of $d_{x^{2}-y^{2}}$-wave altermagnetism is that $\gamma_{1}(\mathbf{Q}_{\textrm{max}})$
does not decrease monotonically as the altermagnetic coupling $\lambda$
increases. Instead, we typically observe the emergence of a local
maximum at relatively large $\lambda$, whose position depends sensitively
on the electron filling. This unexpected enhancement can likely be
attributed to a nesting effect when $\mathbf{Q}$ lies along the diagonal
direction. In general, the optimal magnitude of $\mathbf{Q}$ scales
approximately with the altermagnetic coupling strength $\lambda$,
remaining small for small $\lambda$ and becoming large for large
$\lambda$. However, because of band nesting, both $\mathbf{Q}=(q,q)$
and $\mathbf{\tilde{Q}}=(\pi-q,\pi-q)$ can be equally favorable as
optimal momenta. This indicates that the pairing tendency may persist
even at strong altermagnetic coupling. As a consequence, in the very
low-filling limit, a local maximum of $\gamma_{1}(\mathbf{Q})$ appears
around $\lambda=4t$. As the filling increases, this maximum shifts
progressively toward smaller values of $\lambda$.

In Fig. \ref{fig8}, we present the phase diagram obtained by varying
the altermagnetic coupling strength at a fixed low electron filling
$\nu=0.2$. The critical nearest-neighbor attraction $V_{c}$, required
to drive the system into the superconducting phase, generally decreases
as the altermagnetic coupling increases. As illustrated in the inset,
the dominant pairing channel at weak altermagnetic coupling is the
extended $s$-wave, consistent with the low filling regime. However,
as the coupling strength increases up to $\lambda=2t$, the sub-leading
$p$-wave channel gradually becomes more prominent and eventually
reaches a magnitude comparable to that of the extended $s$-wave component.
As in the case of $d_{xy}$-wave altermagnetism, the superconducting
order parameter in this regime is therefore also multi-component in
character, comprising contributions from both spin-singlet and spin-triplet
channels.

\begin{figure}
\begin{centering}
\includegraphics[width=0.5\textwidth]{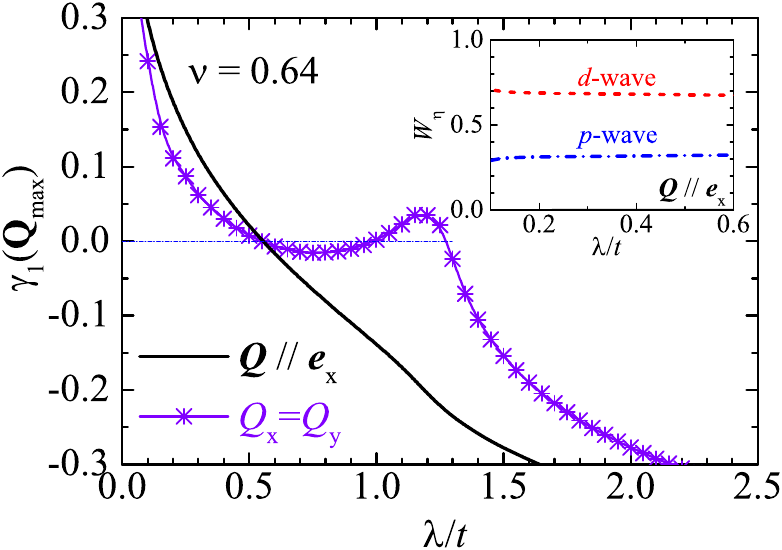}
\par\end{centering}
\caption{\label{fig9} Analysis of the pairing instability under the high-electron
density conditions of Ref. \citep{Jasiewicz2026}. At a large electron
filling factor $\nu=0.64$ and nearest-neighbor interaction $V=-1.6t$,
the maximum of the leading eigenvalue of the inverse vertex function
matrix $\gamma_{1}(\mathbf{Q})$ at small $d_{x^{2}-y^{2}}$-wave
altermagnetic couplings occurs when the center-of-mass momentum $\mathbf{Q}$
is aligned along the $x$-axis. The inset shows how the weights of
the different partial-wave components $W_{\eta}=\left|M_{\eta}\right|^{2}$
vary with the altermagnetic coupling strength $\lambda$.}
\end{figure}

\section{Potential connections to previous work}

Most recently, Jasiewicz \textit{et al.} carried out a mean-field
study of the FFLO superconducting state with a nearest-neighbor attraction
$V=-1.6t$ in the presence of $d_{x^{2}-y^{2}}$-wave altermagnetism
\citep{Jasiewicz2026}. At a relatively high electron filling, $\nu=0.64$,
they identified a singlet-triplet mixed Fulde-Ferrell (FF) order parameter
characterized by a dominant $d$-wave component and a sub-leading
$p$-wave component, emerging when the altermagnetic coupling strength
satisfies $\lambda\gtrsim0.24t$ and the center-of-mass $\mathbf{Q}$
is aligned along the $x$-axis \citep{Jasiewicz2026}. Under identical
conditions, we have performed an analysis of the leading pairing instability.
As shown in Fig. \ref{fig9}, the superconducting transition indeed
sets in for $\lambda<0.55t$ and $\mathbf{Q}$ oriented along the
$x$-axis, and the resulting state contains both $d$-wave and $p$-wave
pairing components (see inset). Furthermore, the ratio between these
two components (approximately $2:1$) is consistent with the relative
magnitudes of the pairing amplitudes reported by Jasiewicz \textit{et
al.} (see Fig. 10(d) of Ref. \citep{Jasiewicz2026}).

\begin{figure}
\begin{centering}
\includegraphics[width=0.5\textwidth]{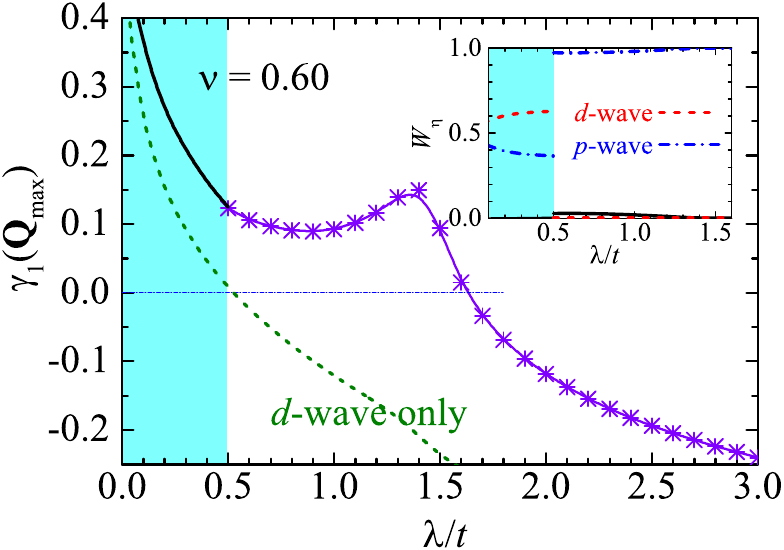}
\par\end{centering}
\caption{\label{fig10} Analysis of the pairing instability under the conditions
of Ref. \citep{Chakraborty2024}, for a filling factor $\nu=0.6$
and nearest-neighbor interaction $V=-2.0t$. $\gamma_{1}(\mathbf{Q}_{\textrm{max}})$
is plotted against the $d_{x^{2}-y^{2}}$-wave altermagnetic coupling
strength $\lambda$, alongside the result obtained by considering
only the $d$-wave pairing channel (green dashed line) as assumed
in Ref. \citep{Chakraborty2024}. The inset illustrates how the weights
of the various partial-wave components $W_{\eta}=\left|M_{\eta}\right|^{2}$
change with $\lambda$.}
\end{figure}

In another theoretical work \citep{Chakraborty2024}, Chakraborty
and Black-Schaffer investigated the FFLO state at the nearest-neighbor
attraction $V=-2t$, again with $d_{x^{2}-y^{2}}$-wave altermagnetism.
Restricting their analysis to the singlet pairing channel and considering
an electron filling of $\nu=0.6$, they found an altermagnetism-driven
FF state within the window $0.44t\leq\lambda\leq0.56t$ \citep{Chakraborty2024}.
In Fig. \ref{fig10}, we have revisited this singlet channel approximation,
by benchmarking it against our full channel calculations. When only
the $d$-wave channel is retained, we obtain a critical altermagnetic
coupling $\lambda_{c}\simeq0.56t$, consistent with the finding by
Chakraborty \textit{et al.} \citep{Chakraborty2024}. However, this
value is substantially smaller than the critical coupling $\lambda_{c}\simeq1.63t$
obtained from the full channel calculation. Actually, as indicated
in the inset, the $p$-wave pairing channel remains significant throughout.
For $\lambda\apprge0.5t$, it becomes nearly the sole dominant pairing
channel.

\section{Considerations on parallel-spin pairing}

\begin{figure}
\begin{centering}
\includegraphics[width=0.5\textwidth]{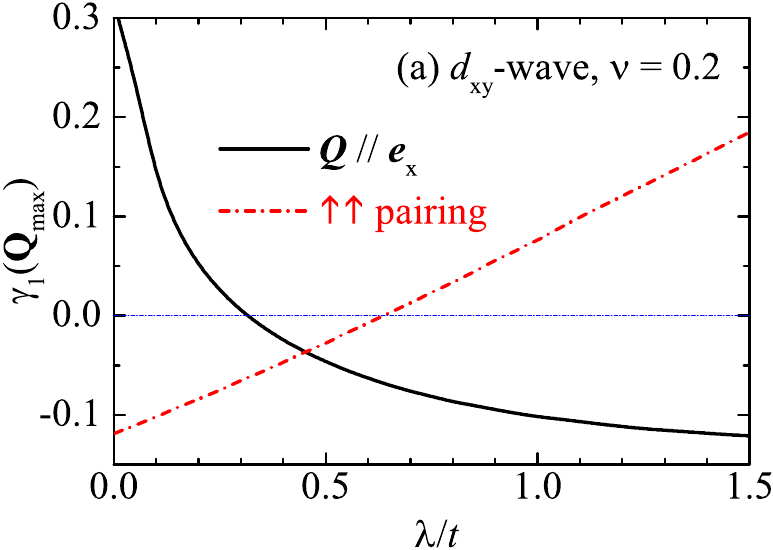}
\par\end{centering}
\begin{centering}
\includegraphics[width=0.5\textwidth]{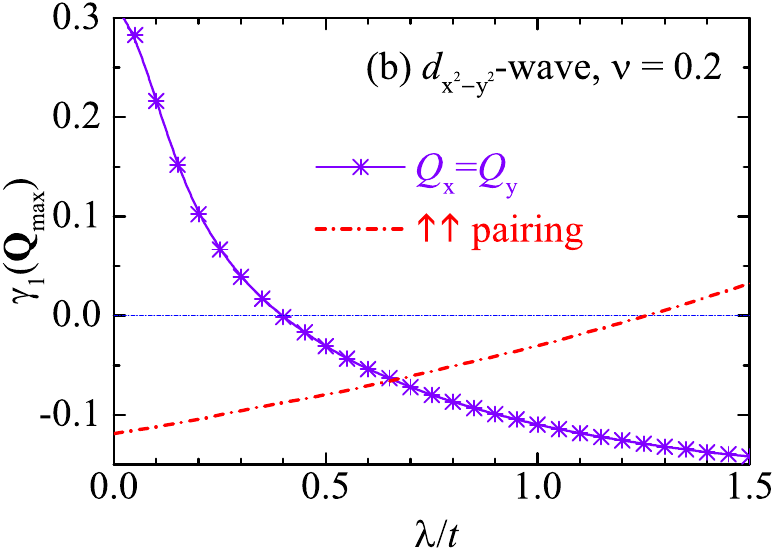}
\par\end{centering}
\caption{\label{fig11} The peak value of the leading eigenvalue of the inverse
vertex function matrix $\gamma_{1}(\mathbf{Q}_{\textrm{max}})$ at
the lattice filling factor $\nu=0.2$ is shown for (a) $d_{xy}$-wave
and (b) $d_{x^{2}-y^{2}}$-wave altermagnetism. Results for the $\uparrow\downarrow$
pairing (black curve in (a) or the line with crosses in (b)) are compared
with those for the triplet $\uparrow\uparrow$ pairing (red dot-dashed
line). Here, we set $V=V_{\sigma}=-1.5t$.}
\end{figure}

Up to this point, we have concentrated on pairing between two electrons
with opposite spins, motivated by the argument illustrated in Fig.
\ref{fig2} that parallel-spin pairing is not the dominant instability
when $\left|V_{\sigma}\right|\leq\left|V\right|$ in the regime of
sufficiently weak altermagnetism. We now examine more carefully the
specific conditions that arise when the nearest-neighbor attractive
interactions have equal strength, $V_{\sigma}=V$.

In Fig. \ref{fig11}, for low electron filling $\nu=0.2$, we compare
the $\lambda$-dependence of $\gamma_{1}(\mathbf{Q}_{\textrm{max}})$
for both anti-parallel-spin pairing and parallel-spin pairing, in
the presence of either $d_{xy}$-wave (a) or $d_{x^{2}-y^{2}}$-wave
(b) altermagnetism. For parallel-spin pairing, the instability always
occurs at $\mathbf{Q}=0$ (not shown), regardless of whether altermagnetism
is present. Moreover, as the altermagnetic coupling $\lambda$ increases,
the leading eigenvalue $\gamma_{1}(\mathbf{Q}_{\textrm{max}})$ tends
to increases rather than decreases. Consequently, at sufficiently
large altermagnetic coupling, the dominant pairing instability eventually
shifts from anti-parallel-spin pairing to parallel-spin pairing. However,
for altermagnetic coupling below the critical value $\lambda_{c}$,
the leading pairing instability consistently arises from electrons
with anti-parallel spins. Therefore, the phase diagrams shown in Fig.
\ref{fig6}(a) and Fig. \ref{fig8} remain valid in the regime of
weak altermagnetic coupling.

\begin{figure}
\begin{centering}
\includegraphics[width=0.5\textwidth]{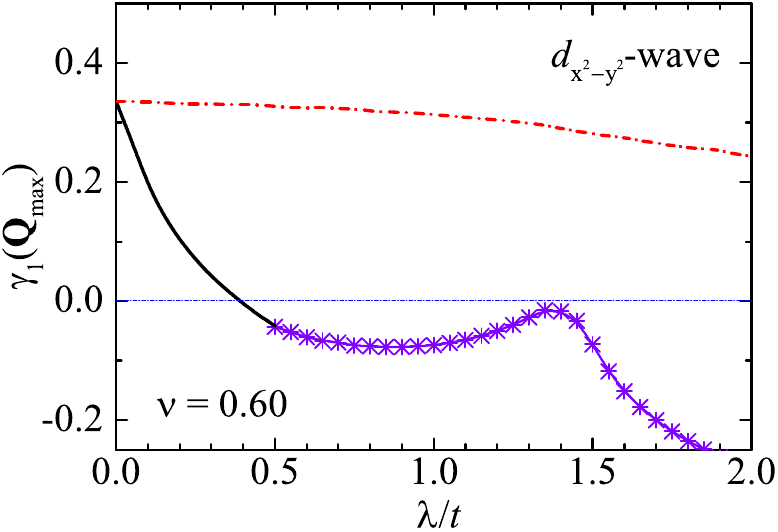}
\par\end{centering}
\caption{\label{fig12} The same as in Fig. \ref{fig11}(b), except evaluated
for a lattice filling factor of $\nu=0.6$.}
\end{figure}

This behavior changes at higher electron fillings. As illustrated
in Fig. \ref{fig12}, taking $d_{x^{2}-y^{2}}$-wave altermagnetism
as a representative example, at electron filling $\nu=0.6$ the leading
pairing instability now occurs among electrons with same spins at
all altermagnetic couplings. This parallel-spin, triplet pairing state
has recently been explored in the literature, both in the presence
\citep{Zhu2023} and absence \citep{Hong2025} of spin--orbit coupling.

\section{Conclusions and outlooks}

In summary, we have investigated the pairing instability in an altermagnetic
metal subject to a finite-range attractive interaction that accommodates
multiple pairing channels with distinct parity and symmetry. The system
is described by an altermagnetic Hubbard $UV$ model incorporating
both on-site ($U$) and nearest-neighbor ($V$) attractive interactions.
Within the standard non-self-consistent many-body $T$-matrix approximation
\citep{MahanManyParticlePhysics}, we have determined the two-particle
vertex function by projecting it onto the different pairing channels.
By subsequently diagonalizing the resulting inverse vertex matrix,
we have identified the leading pairing instability according to the
well-established Thouless criterion \citep{Thouless1960,Liu2006}.

In this way, we have confirmed the emergence of altermagnetism-driven
FFLO superconducting states, which have already been examined in some
detail in earlier works \citep{Sumita2023,Zhang2024,Chakraborty2024,Hong2025,Sumita2025,Hu2025PRB,Liu2026PRB}.
Our systematic analysis of the pairing instability, however, goes
beyond the commonly adopted single-pairing-channel approximation and
demonstrates that finite-momentum pairing naturally leads to the admixture
of multiple pairing channels with different parity. Consequently,
altermagnetic FFLO states generally exhibit a multi-component order
parameter containing both spin-singlet and spin-triplet contributions
\citep{Jasiewicz2026}. At low electron filling, this result is consistent
with the previous exact two-electron calculation \citep{Hu2026},
which predicts finite-momentum bound pairs with mixed singlet-triplet
character.

We emphasize that the singlet-triplet mixing discussed here arises
in a different physical setting from the admixture commonly considered
in non-centrosymmetric superconductors \citep{Smidman2017}. In the
latter, the absence of inversion symmetry permits parity mixing through
antisymmetric spin-orbit coupling, resulting in superconducting states
with coexisting even-parity singlet and odd-parity triplet components.
In contrast, the mixing considered here originates from the momentum-dependent
spin splitting associated with altermagnetic order and is intrinsically
linked to finite-momentum pairing. Thus, although both situations
can give rise to singlet-triplet admixture, the mixing discussed here
is a consequence of the momentum-dependent spin texture and symmetry
of the altermagnetic state, rather than parity mixing arising from
the absence of inversion symmetry.

It would be of considerable interest to investigate these predictions
in altermagnetic superconductors, whose superconducting state remains
to be experimentally established. In a multi-orbital Hubbard-Kanamori
description, repulsive intra- and inter-orbital Coulomb interactions,
together with Hund\textquoteright s coupling, can enhance spin fluctuations
and promote spin-density-wave (SDW) instabilities. The favored ordering
wavevector and symmetry depend on the orbital-resolved band structure
and spin susceptibility; possible form factors in a tetragonal system
include $d_{x^{2}-y^{2}}$- and $d_{xy}$-like structures. Although
an SDW-type instability may underlie altermagnetic order, not every
SDW state is altermagnetic. Once altermagnetism is established, its
momentum-dependent spin splitting can further modify the particle-hole
susceptibilities \citep{Parthenios2025}. Charge-density-wave (CDW)
instabilities may also be favored by nonlocal interactions, electron-phonon
coupling, or favorable Fermi-surface features; repulsive onsite interactions
alone do not generically imply an attractive charge channel. Meanwhile,
effective attractive interactions between neighboring sites may emerge
from onsite Coulomb repulsion mediated by spin fluctuations \citep{Nakajima1973,Hirsch1985,Scalapino1986,Ohashi1993,Romer2015,Wu2025}.
Determining the leading SDW or CDW channel and its competition with
superconductivity requires a microscopic multi-orbital analysis of
the particle-hole susceptibilities and interaction vertices, which
we leave for future work.

From the perspective of symmetry, several layered tetragonal materials
may provide promising platforms. In particular, metallic KV$_{2}$Se$_{2}$O
and RbV$_{2}$Te$_{2}$O have been reported to exhibit $d$-wave altermagnetic
spin splitting and quasi-two-dimensional electronic structures \citep{Jiang2025,Zhang2025}.
RuO$_{2}$ has also been extensively discussed as a prototypical $d$-wave
altermagnetic candidate \citep{Ahn2019}, although the microscopic
origin and bulk magnetic state of RuO$_{2}$ remain under active investigation.

We also note that a more realistic description of these potential
altermagnetic systems may require a refined model Hamiltonian incorporating
sub-lattice degrees of freedom \citep{Roig2024,Sumita2025} and higher
even-parity altermagnetic orders (e.g., $g$- or $i$-wave). The resulting
multiband electronic structure gives rise to multiple Fermi surfaces,
which may substantially modify the FFLO state through interband coupling,
band-dependent pairing, and Fermi-surface mismatch. Consequently,
the stability, modulation wave vector, and the associated singlet-triplet
mixing of the FFLO state may be significantly affected. A quantitative
investigation of these multiband effects remains an important direction
for future work.

Another interesting direction is to explore the continuum limit of
a two-dimensional electron gas. In the dilute regime, the kinetic
Hamiltonians corresponding to $d_{xy}$- and $d_{x^{2}-y^{2}}$-wave
altermagnetism are related by a $45^{\circ}$ rotation. The electron-electron
interaction can be characterized by scattering lengths in different
partial-wave channels. For the simplest $s$-wave case, a detailed
analysis has been presented in Ref. \citep{Wang2026}. Motivated by
our results at low electron fillings, we anticipate that the FFLO
superconducting state and the associated singlet-triplet mixing may
also emerge in this continuum setting.

Finally, in the mixed singlet-triplet superconducnting phase, it is
also important to investigate the role of relative phases between
different components of the order parameter. The breaking of time-reversal
symmetry can give rise to spontaneous surface currents, similar to
those found in topological insulators, and this effect has already
been discussed for the $s+id_{xy}$ spin-singlet state \citep{Kusama1999}.
Extending the analysis from multi-component singlet-singlet pairing
to the singlet-triplet case presents an interesting and worthwhile
direction for further study.
\begin{acknowledgments}
This research was supported by the Australian Research Council's (ARC)
Discovery Program, Grants Nos. DP240101590 (H.H.), FT230100229 (J.W.)
and DP240100248 (X.-J.L.). Y.O. was supported by JSPS KAKENHI Grant
Numbers JP22K03486, JP26K06983.
\end{acknowledgments}

\appendix

\section{Numerical calculation of the pair propagator matrix}

\begin{figure}
\begin{centering}
\includegraphics[width=0.5\textwidth]{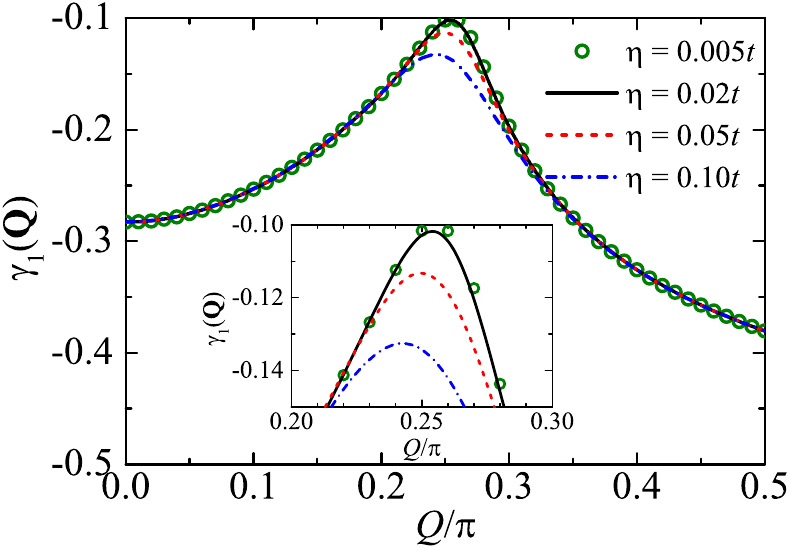}
\par\end{centering}
\caption{\label{fig13} The dependence of the largest eigenvalue of the inverse
vertex function matrix $\gamma_{1}(\mathbf{Q})$ on the spectral broadening
factor $\eta$: $\eta=0.005t$ (circles), $\eta=0.02t$ (black solid
line), $\eta=0.05t$ (red dashed line) and $\eta=0.10t$ (blue dot-dashed
line). The inset zooms in on the region near the peak. Results are
shown for $d_{xy}$-wave altermagnetism with coupling strength $\lambda=1.0t$
and filling factor $\nu=0.2$. Other parameters are $U=-0.01t$ and
$V=-1.5t$. }
\end{figure}

In evaluating the pair propagator matrix Eq. (\ref{eq:PairPropagator}),
it is necessary to eliminate the sharp structure in the integrand
that arises when the energy $\omega$ lies within the two-electron
continuum. To address this issue, we introduce a small spectral broadening
factor $\eta>0$ by shifting the energy as, $\omega\rightarrow\omega+i\eta$.
We then extrapolate to the zero-broadening limit $\eta=0^{+}$ by
using a cubic polynomial fitting. 

Fig. \ref{fig13} presents an example for $d_{xy}$-wave altermagnetism
with coupling strength $\lambda=1.0t$ and low electron filling $\nu=0.2$.
The leading eigenvalue of the inverse vertex function matrix, $\gamma_{1}(\mathbf{\mathbf{Q}}=Q\mathbf{e}_{x})$,
is shown for several values of the spectral broadening factor $\eta$,
as indicated. Although the cubic fitting procedure does not completely
eliminate the $\eta$-dependence, the remaining dependence becomes
negligible for sufficiently small $\eta$. Throughout this work, we
typically take $\eta=0.02t$. The estimated absolute error in the
leading eigenvalue is on the order of $10^{-3}t^{-1}$, which is barely
discernible in the figure.

\begin{figure}
\begin{centering}
\includegraphics[width=0.5\textwidth]{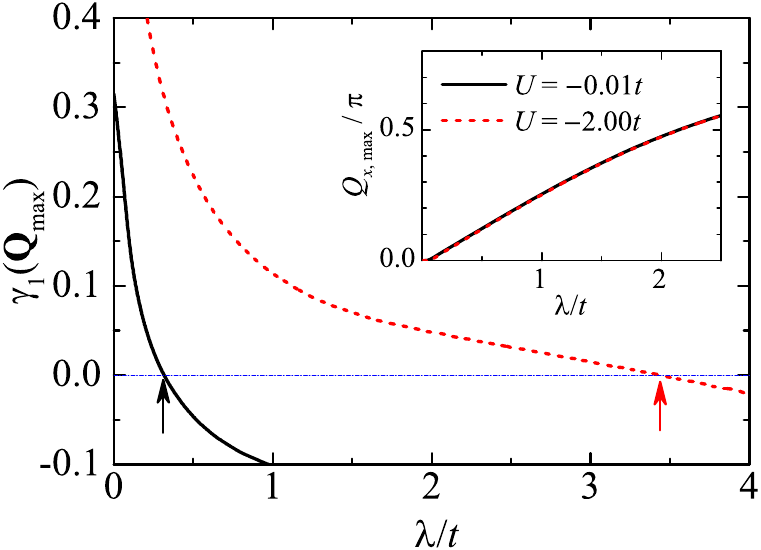}
\par\end{centering}
\caption{\label{fig14} The maximum value of the leading eigenvalue of the
inverse vertex function matrix $\gamma_{1}(\mathbf{Q}_{\textrm{max}})$
is shown as a function of the $d_{xy}$-wave altermagnetic coupling
strength $\lambda$ for two different on-site attractive interactions:
$U=-0.01t$ (black curve) and $U=-2.0t$ (red dashed line). The inset
shows the magnitude of the center-of-mass momentum $\mathbf{Q}$ at
which this maximum occurs, with $\mathbf{Q}$ oriented along the $x$-axis.
The nearest-neighbor interaction strength is $V=-1.5t$ and the electron
filling is $\nu=0.2$.}
\end{figure}

\begin{figure}
\begin{centering}
\includegraphics[width=0.5\textwidth]{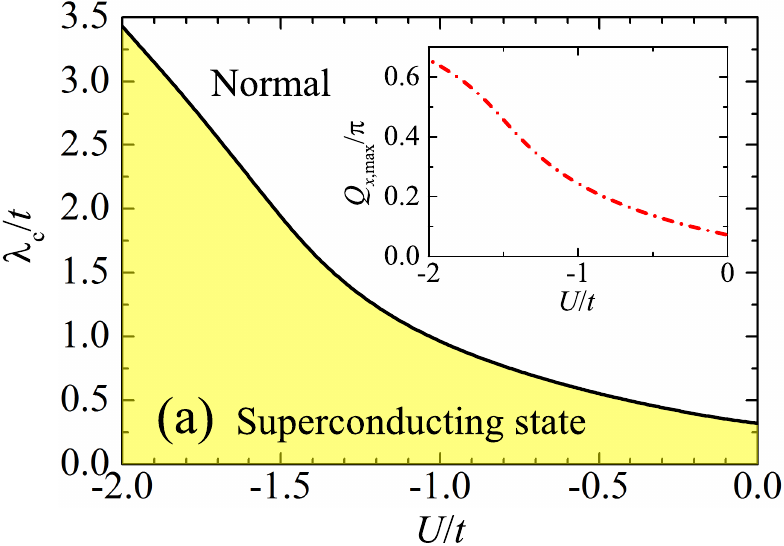}
\par\end{centering}
\begin{centering}
\includegraphics[width=0.5\textwidth]{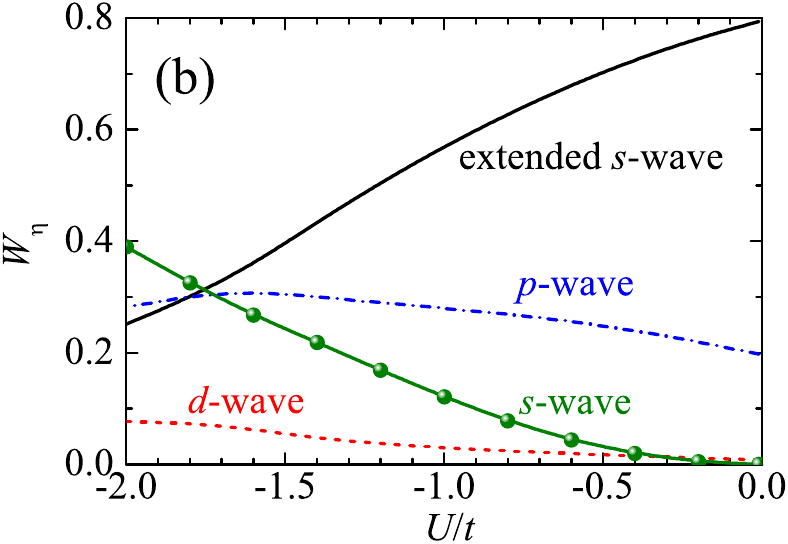}
\par\end{centering}
\caption{\label{fig15} (a) Phase diagram for $d_{xy}$-wave altermagnetism
as a function of on-site attraction: the critical altermagnetic coupling
strength $\lambda_{c}$ is plotted against the on-site attractive
interaction strength $U$. The inset depicts the center-of-mass momentum
$\mathbf{Q}=Q\mathbf{e}_{x}$ at the superconducting phase boundary.
(b) The contributions of different partial-wave components to the
dominant pairing channel $W_{\eta}=\left|M_{\eta}\right|^{2}$ at
the superconducting transition. The nearest-neighbor interaction strength
is $V=-1.5t$ and the electron filling is $\nu=0.2$.}
\end{figure}

\section{The effect of on-site attractive interactions}

Here, we briefly discuss how the pairing instability is modified by
incorporating an on-site attractive interaction. In Fig. \ref{fig14},
we report $\gamma_{1}(\mathbf{Q}_{\textrm{max}})$ both in the absence
and presence of on-site attraction, under $d_{xy}$-wave altermagnetism.
As shown in the inset, the peak center-of-mass momentum $\mathbf{Q}$
remains nearly independent of the strength of the on-site attraction.
In contrast, the critical altermagnetic coupling strength $\lambda_{c}$
increases substantially when the on-site attraction is included, as
indicated by the red arrow in the figure.

In Fig. \ref{fig15}(a), we present the resulting critical altermagnetic
coupling $\lambda_{c}$ as a function of the on-site attraction, which
defines the corresponding phase diagram. Fig. \ref{fig15}(b) further
displays the contributions of different partial-wave components to
the emerging superconducting order parameter at the phase transition.
As the on-site attraction becomes stronger, the purely $s$-wave component
is progressively enhanced. At about $U\simeq-1.75t\sim V$, it exceeds
the extended $s$-wave contribution and eventually becomes the dominant
pairing channel, as expected.

\end{document}